\documentclass[twocolumn,times,tighten]{aastex631}

\usepackage{xspace}
\usepackage[T1]{fontenc}

\newcommand{\FeKa}{Fe K\ensuremath{\alpha}\xspace}

\newcommand{\NH}{\ensuremath{N_{\mathrm{H}}}\xspace}

\newcommand{\xmm}{{XMM-Newton}\xspace}
\newcommand{\chandra}{{Chandra}\xspace}
\newcommand{\swift}{{Swift}\xspace}

\newcommand{\hst}{{HST}\xspace}

\newcommand{\nustar}{{NuSTAR}\xspace}

\newcommand{\cm}{{\ensuremath{\rm{cm}^{-2}}}\xspace}

\newcommand{\lya}{Ly\ensuremath{\alpha}\xspace}

\newcommand{\civ}{\ion{C}{4}\xspace}
\newcommand{\cii}{\ion{C}{2}\xspace}

\newcommand{\ngc}{{NGC~5548}\xspace}

\newcommand{\cx}{\ensuremath{C_{\rm X}}\xspace}
\newcommand{\cuv}{\ensuremath{C_{\rm UV}}\xspace}

\newcommand{\ewciv}{EW\ensuremath{_{\rm C\, IV}}\xspace}

\usepackage{xcolor}

\usepackage{xfrac}
\received{2022 June 1}
\accepted{2022 July 19 by ApJL}
\shorttitle{10-Year Transformation of the Obscuring Wind in NGC 5548}
\shortauthors{Mehdipour et al.}
\graphicspath{{./}{figures/}}
\begin{document}

\title{\vspace{-0.2cm} \Large 10-Year Transformation of the Obscuring Wind in NGC 5548}

\author[0000-0002-4992-4664]{Missagh Mehdipour}
\affiliation{Space Telescope Science Institute, 3700 San Martin Drive, Baltimore, MD 21218, USA; \href{mailto:mmehdipour@stsci.edu}{mmehdipour@stsci.edu}}

\author[0000-0002-2180-8266]{Gerard A. Kriss}
\affiliation{Space Telescope Science Institute, 3700 San Martin Drive, Baltimore, MD 21218, USA; \href{mailto:mmehdipour@stsci.edu}{mmehdipour@stsci.edu}}

\author[0000-0001-8470-749X]{Elisa Costantini}
\affiliation{SRON Netherlands Institute for Space Research, Niels Bohrweg 4, 2333 CA Leiden, the Netherlands}
\affiliation{Anton Pannekoek Institute, University of Amsterdam, Postbus 94249, 1090 GE Amsterdam, The Netherlands}

\author[0000-0001-9911-7038]{Liyi Gu}
\affiliation{SRON Netherlands Institute for Space Research, Niels Bohrweg 4, 2333 CA Leiden, the Netherlands}
\affiliation{RIKEN High Energy Astrophysics Laboratory, 2-1 Hirosawa, Wako, Saitama 351-0198, Japan}
\affiliation{Leiden Observatory, Leiden University, PO Box 9513, 2300 RA Leiden, the Netherlands}

\author[0000-0001-5540-2822]{Jelle S. Kaastra}
\affiliation{SRON Netherlands Institute for Space Research, Niels Bohrweg 4, 2333 CA Leiden, the Netherlands}
\affiliation{Leiden Observatory, Leiden University, PO Box 9513, 2300 RA Leiden, the Netherlands}

\author{Hermine Landt}
\affiliation{Centre for Extragalactic Astronomy, Department of Physics, Durham University, South Road, Durham DH1 3LE, UK}

\author[0000-0001-7557-9713]{Junjie Mao}
\affiliation{Department of Astronomy, Tsinghua University, Haidian DS 100084, Beijing, People’s Republic of China}
\affiliation{SRON Netherlands Institute for Space Research, Niels Bohrweg 4, 2333 CA Leiden, the Netherlands}

\begin{abstract}
A decade ago the archetypal Seyfert-1 galaxy \ngc was discovered to have undergone major spectral changes. 
The soft X-ray flux had dropped by a factor of 30 while new broad and blueshifted UV absorption lines appeared. 
This was explained by the emergence of a new obscuring wind from the accretion disk.
Here we report on the striking long-term variability of the obscuring disk wind in \ngc including new observations taken in 2021--2022 with the \swift Observatory and the Hubble Space Telescope's (HST) Cosmic Origins Spectrograph (COS).
The X-ray spectral hardening as a result of obscuration has declined over the years, reaching its lowest in 2022 at which point we find the broad \civ UV absorption line to be nearly vanished.
The associated narrow low-ionization UV absorption lines, produced previously when shielded from the X-rays, are also remarkably diminished in 2022.
We find a highly significant correlation between the variabilities of the X-ray hardening and the equivalent width of the broad \civ absorption line, demonstrating that X-ray obscuration is inherently linked to disk winds.
We derive for the first time a relation between the X-ray and UV covering fractions of the obscuring wind using its long-term evolution.
The diminished X-ray obscuration and UV absorption are likely caused by an increasingly intermittent supply of outflowing streams from the accretion disk.   
This results in growing gaps and interstices in the clumpy disk wind, thereby reducing its covering fractions.
\end{abstract}
\section{Introduction} 
\label{sect_intro}
Winds in active galactic nuclei (AGNs) bridge the supermassive black holes (SMBHs) to their environment. They transport mass and energy away from the central engine and into the host galaxy of the AGN. This injection of momentum and energy can have important implications for the co-evolution of SMBHs and galaxies through the resulting feedback mechanism between the AGN activity and star formation (e.g. \citealt{King15,Gasp17,Harr18}). Thus, ascertaining the physical properties and energetics of different types of AGN winds, and understanding how they are launched and driven, are important for determining their role and impact in AGN feedback.

Bright Seyfert-1 galaxies are useful laboratories for ascertaining the physical structure of AGN winds. Most notably \ngc has been at the forefront of discoveries. The first X-ray absorption lines of `warm-absorber outflows' were found in \ngc with \chandra soon after its launch \citep{Kaas00}. At that epoch this archetypal Seyfert-1 galaxy was in its usual X-ray bright state. Thirteen years later in 2013 a large multi-wavelength campaign, including \xmm, \nustar, \swift, and \hst was performed on \ngc for a comprehensive mapping of its warm absorber. However, unexpectedly, \ngc was discovered to have become obscured in the X-rays \citep{Kaas14,Meh15a} while new UV absorption lines, such as broad \civ and narrow \cii lines, were found in the HST/COS spectra \citep{Arav15,Kris19b}. Also, pronounced soft X-ray emission lines stood out above the obscured X-ray continuum of \ngc \citep{Kaas14,Mao18}, reminiscent of the X-ray spectra of Seyfert-2 galaxies, such as NGC~1068 \citep{Graf21}. The studies of this new state of \ngc showed that the central X-ray source is being obscured by a massive stream of outflowing gas in the vicinity of the accretion disk, extending to and beyond the broad-line region (BLR). This is referred to as an `obscuring disk wind', or `obscurer' for short.

While obscuration is generally associated with type-2 AGN, and broad-absorption lines (BALs) are typically found in powerful BAL quasars, \ngc showed their presence together in a Seyfert-1 galaxy. Interestingly, in quasars such shielding of the UV-absorbing gas from the ionizing X-ray source is required in order to facilitate radiative driving of their outflows \citep{Prog04}. However, the potential role of this X-ray shielding in radiation-driven winds remains debatable in the literature: the hydrodynamical simulation study of \citet{Higg14} finds that the shielding may not be effective in keeping the outer UV absorbers from being over-ionized because of reprocessing and scattering of the ionizing X-rays; and the study of BAL quasars by \citet{Luo14} suggests that they are intrinsically X-ray faint and do not need shielding to drive winds.

The obscuring wind in \ngc shields the warm-absorber outflows that are located further out in the torus region and the narrow-line region (NLR), causing them to become less ionized \citep{Arav15,Kris19b}. This is evident by the appearance of new low-ionization narrow absorption lines, such as \ion{C}{2} in the UV \citep{Arav15} and \ion{He}{1} in the infrared \citep{Wild21}. The effects of obscuration in \ngc is not limited only to the line-of-sight absorption. Extraordinary periods of de-correlation between the variabilities of the UV continuum and the BLR emission lines were found in \ngc \citep{Goad16}. This has been explained as a consequence of global obscuration in all directions, shielding the BLR from the ionizing source \citep{Dehg19b}. Thus, the obscurer in \ngc alters the ionization state of both the warm-absorber outflows \citep{Dehg19a} and the BLR \citep{Dehg19b}, impacting our interpretation of their characteristics and variability. The transient nature of the obscuring disk wind in the BLR and the persistent nature of the warm absorber in the NLR suggests that these outflows are not physically connected at the epoch of the observations. However, the origin and formation history of the warm absorbers, and whether or not they are disk winds, are still open questions.

The Swift Observatory has played an instrumental role in the studies of obscuring winds in \ngc \citep{Mehd16a} and other AGN. \swift monitoring searches have led to target-of-opportunity observations with \xmm, \nustar, and HST in additional Seyfert-1 galaxies, including NGC~3783 \citep{Mehd17,Kris19a} and NGC~3227 \citep{Mehd21,Mao22}. There are however puzzling differences between the characteristics of the obscuring winds in \ngc and other AGN. The obscurer in \ngc has been continuously present for one decade: since at least February 2012 in \swift data \citep{Mehd16a} and June 2011 in HST/COS data \citep{Kaas14}. However, in other targets the obscuration is seen to be a transient event lasting weeks/months \citep{Kaas18}. It is currently uncertain what physical factors govern the occurrence and the duty cycle of the obscuration events. Also, the UV absorption counterpart of the X-ray obscuration is not always detected, such as in NGC~3227 \citep{Mao22}. It is unclear whether this is due to ionization or geometrical effects. Thus, a general understanding of the relation between X-ray obscuration and UV absorption is yet to be achieved. \ngc, with its extensive database and comprehensive studies, is a pivotal target for advancing further our understanding of the obscuring winds in AGN.

\begin{figure}
\centering
\resizebox{1.0\hsize}{!}{\includegraphics[angle=0]{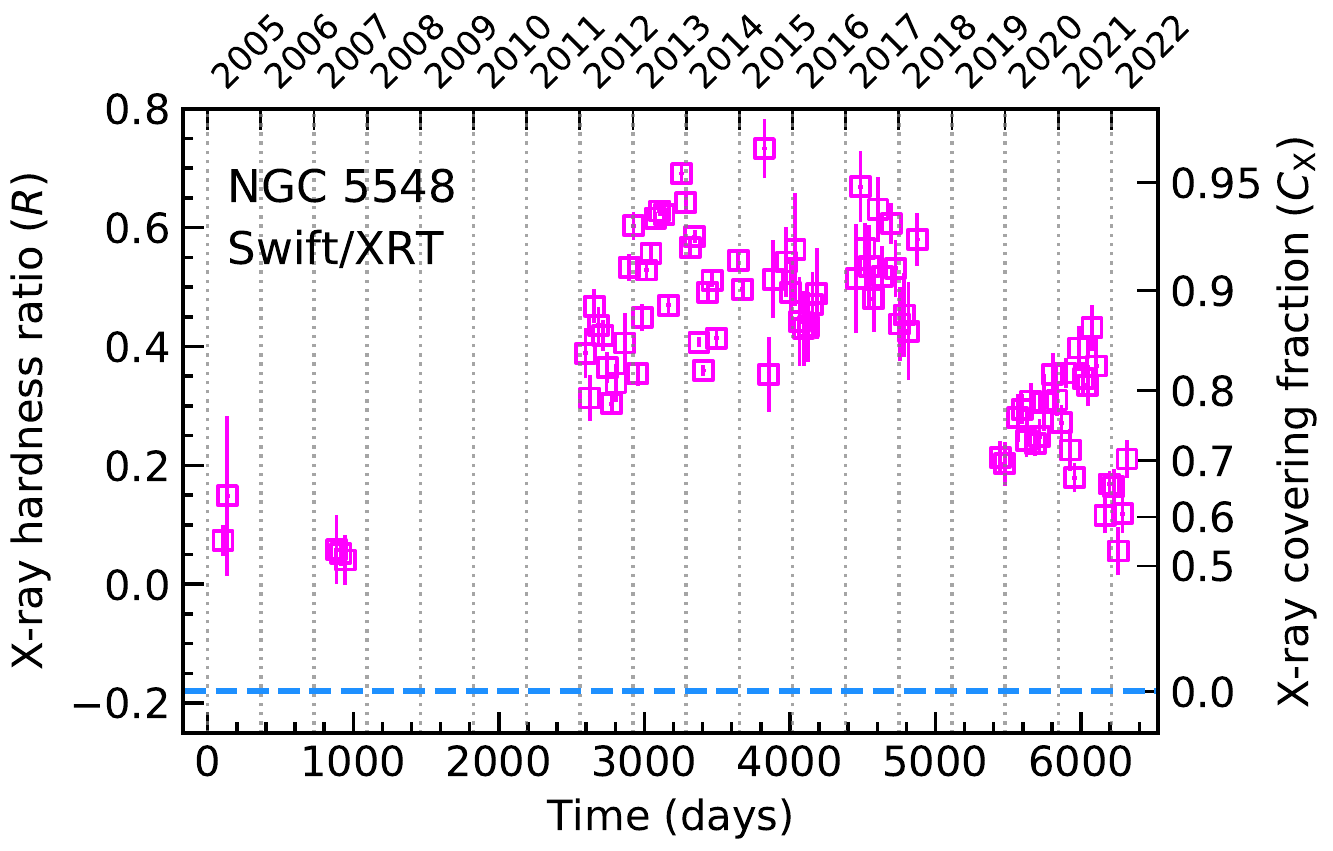}}
\caption{\swift/XRT hardness ratio ($R$) lightcurve of \ngc. It spans from April 2005 to April 2022 with a time binning of one month. Vertical dotted lines are drawn at first day of each calender year for reference. For data since 2012 the corresponding X-ray covering fraction (\cx) of the obscuring wind is shown on the right axis. The 2005 and 2007 data are not obscured (see Sect. \ref{sect_model}). The horizontal dashed line is drawn at $R$ with zero obscuration (${\cx=0}$) derived from the 2002 unobscured \chandra spectrum. Long-term rise and decline of the obscuration are evident.
\label{fig_lc}}
\vspace{-0.1cm}
\end{figure}

A key measurement from \swift monitoring is the X-ray hardness ratio ($R$), defined as ${R = (H - S)/(H + S)}$, where $H$ and $S$ are the \swift X-ray Telescope (XRT) count rates in the hard (1.5--10 keV) and soft (0.3--1.5 keV) bands, respectively. The hardness ratio $R$ is a useful tracer of X-ray obscuration. The study of \citet{Mehd16a} showed that $R$ variability in \ngc is directly related to changes in the X-ray covering fraction (\cx) of the obscurer. Figure \ref{fig_lc} shows the \swift $R$ lightcurve of \ngc spanning from its first observation in April 2005 to the latest one in April 2022. The data suggest long-term obscuration changes, declining in recent years. To further investigate the disk wind evolution and ascertain the relation between its X-ray obscuration and UV absorption, we obtained new spectra in 2021--2022, which are described in the following section.

%

\section{New observations} 
\label{sect_data}
%

%
\begin{figure}
\centering
\resizebox{0.96\hsize}{!}{\includegraphics[angle=270]{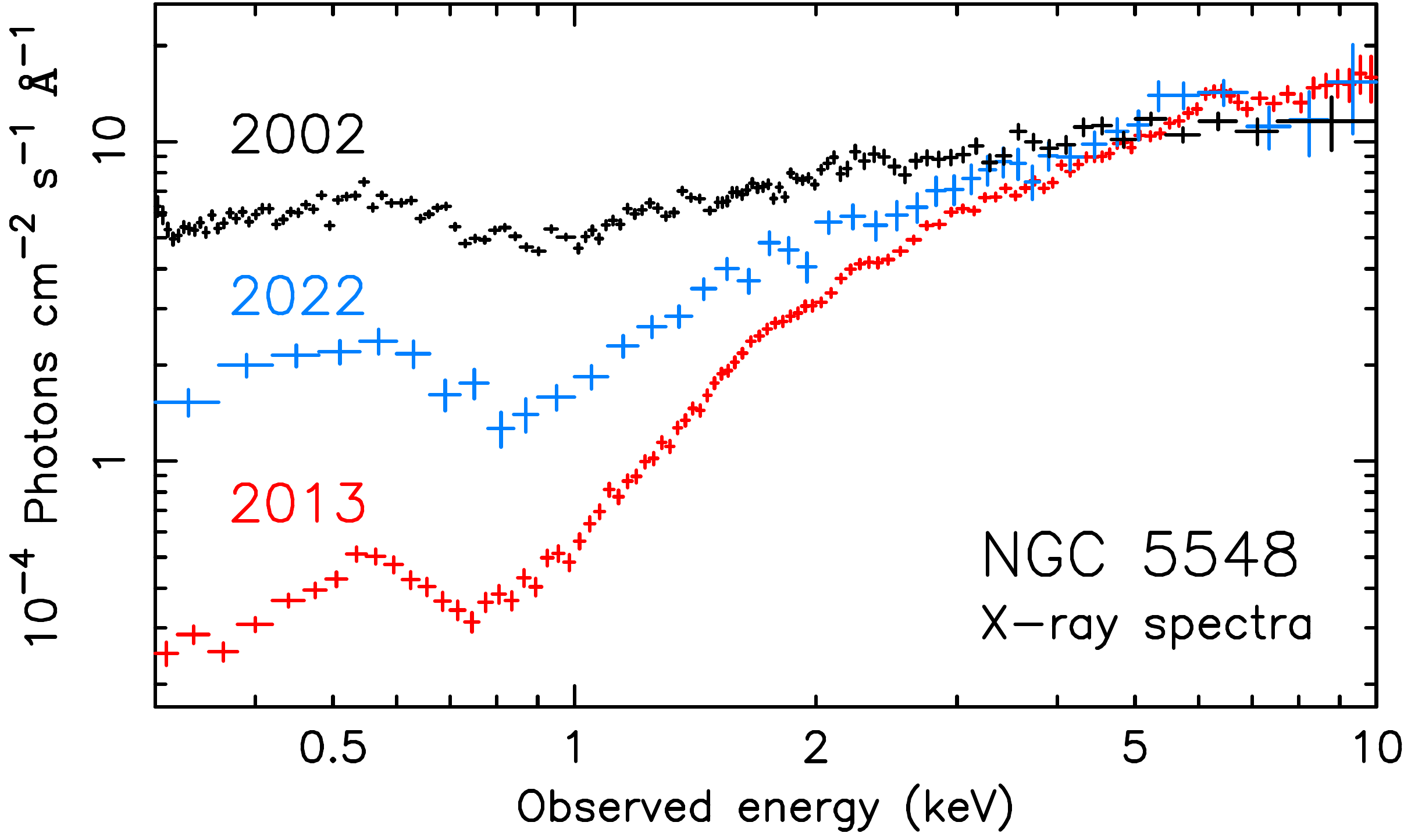}}
\resizebox{0.94\hsize}{!}{\includegraphics[angle=0]{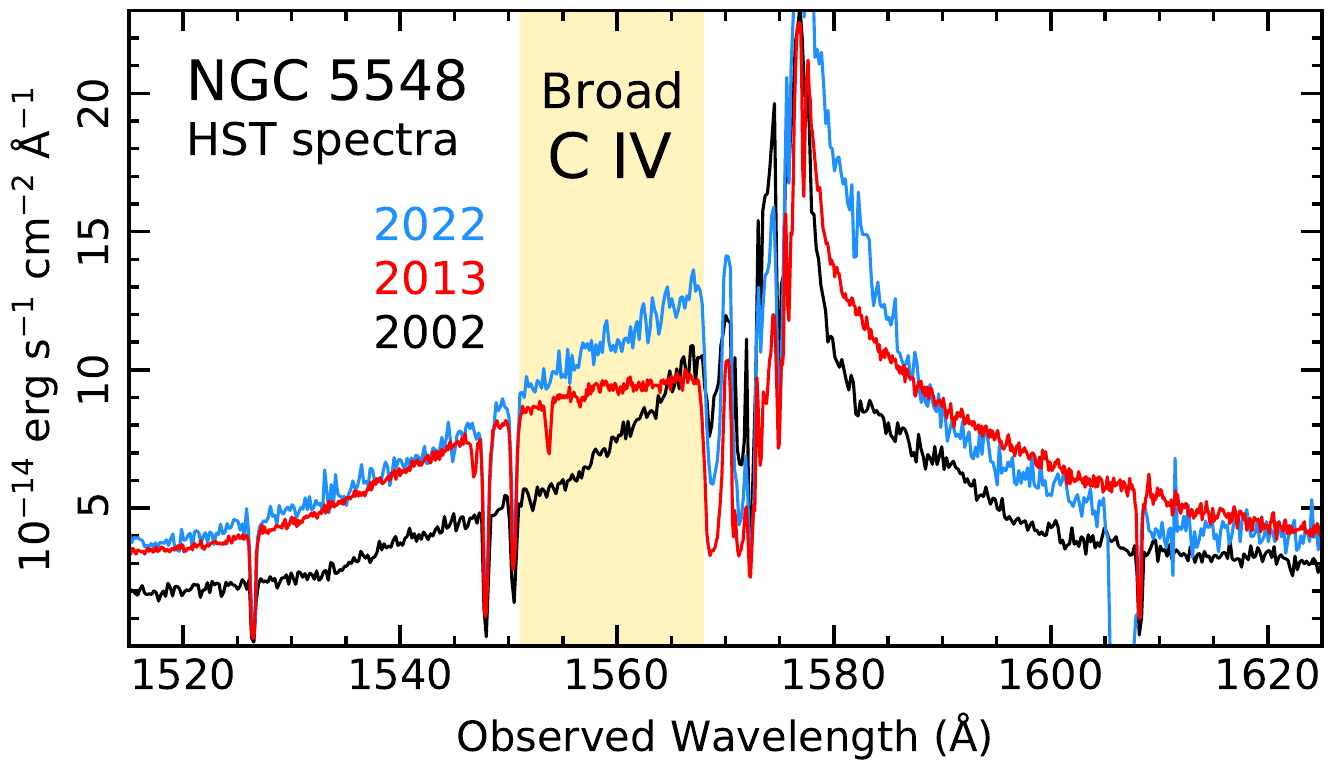}}
\resizebox{0.94\hsize}{!}{\includegraphics[angle=0]{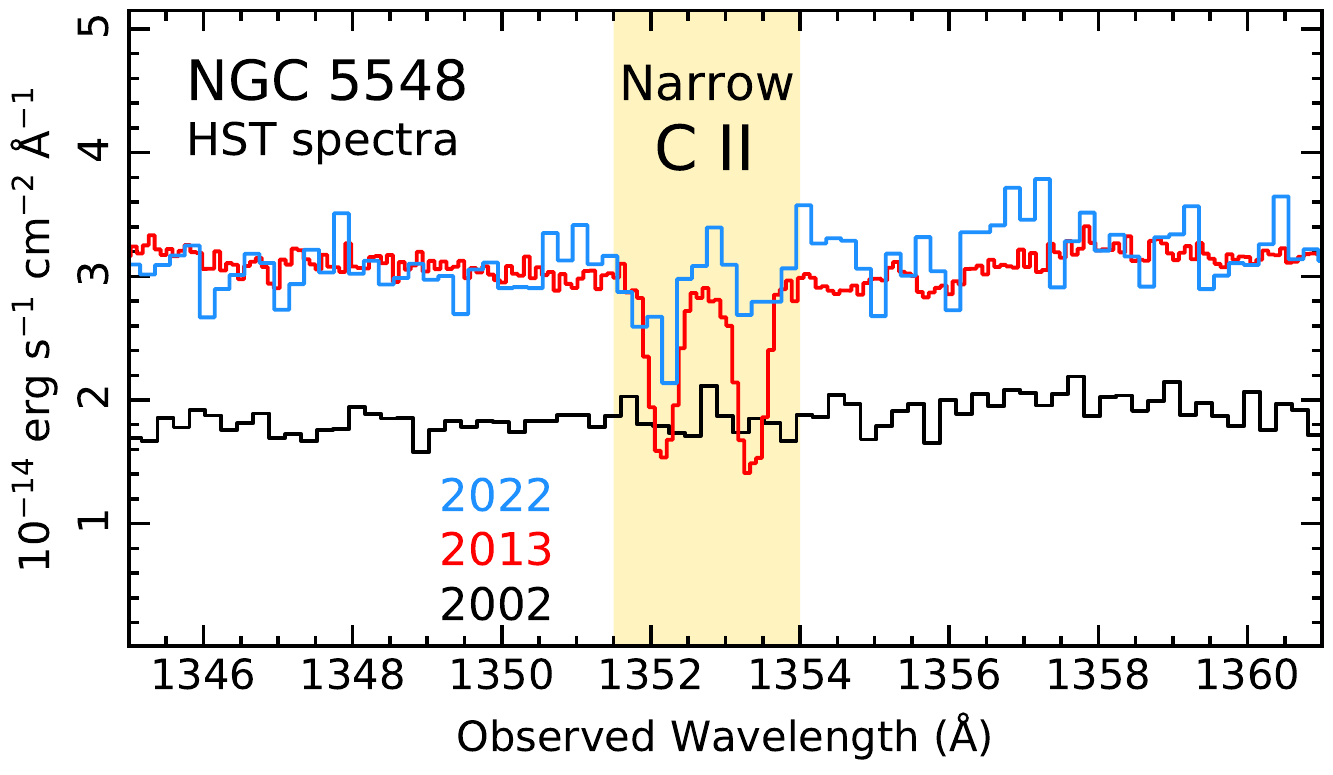}}
\caption{X-ray (top panel) and UV (middle and bottom panels) spectra of \ngc taken at epochs of zero obscuration (2002), strong obscuration (2013), and diminished obscuration (2022).
The 2002 spectra are from \chandra/LETG and HST/STIS. The 2013 and 2022 spectra are from \swift/XRT and HST/COS. The UV spectral regions of interest are highlighted in yellow. In 2022 the X-ray obscuration (top panel), the broad \ion{C}{4} absorption (middle panel), and the narrow \ion{C}{2} absorption (bottom panel) are diminished together.  
\label{fig_spec}}
\vspace{0.2cm}
\end{figure}

Our new HST/COS observations of \ngc (ID: 16842) were taken on 2021-12-30 (1 orbit) and 2022-01-31 (1 orbit). The data can be accessed via \dataset[10.17909/4p21-1m98]{\doi{10.17909/4p21-1m98}}. Each visit used the G130M and G160M gratings to measure the \lya and \civ lines. The spectra from the two visits are very similar. The COS data were processed with the latest calibration pipeline, CalCOS v3.3.11. The \swift/XRT data products of \ngc are obtained from the UK Swift Science Data Centre tools \citep{Evan07,Eva09} and prepared for modeling according to \citet{Mehd16a}. The 2022 \swift/XRT data (15 ks) were used for producing a time-averaged X-ray spectrum for the current epoch. Also, new \chandra High Energy Transmission Grating (HETG) observations were taken for the spectroscopy of the \FeKa line and the warm absorber in the hard X-ray band, which will be presented separately in a future paper. For measuring changes in the obscurer, which is the aim of this paper, the \swift/XRT spectrum is more suitable than HETG as it covers the full 0.3--10 keV band and has a higher signal-to-noise.

For comparing the new 2022 spectra with those from epochs of zero obscuration (2002) and strong obscuration (2013), as shown in Figure \ref{fig_spec}, we retrieved the following archival spectra: the 2002 HST Space Telescope Imaging Spectrograph (STIS) spectrum, the 2002 \chandra Low Energy Transmission Grating (LETG) spectrum, the 2013 HST/COS spectrum, and the 2013 \swift/XRT spectrum. These archival STIS, COS, and LETG spectra are taken from \citet{Kaas14}, and the time-averaged 2013 XRT spectrum is produced following the procedure described in \citet{Mehd16a}.

\section{Modeling of the obscuring wind} 
\label{sect_model}

In order to investigate any new changes in the parameters of the obscuring wind, we apply the spectral models that we derived in our previous X-ray \citep{Mehd16a} and UV \citep{Kris19b} studies of the obscurer in \ngc. This enables a uniform and consistent comparison of the parameters' variabilities, allowing us to examine the long-term evolution of the obscuring wind. The study of \citet{Mehd16a} incorporated the continuum spectral energy distribution (SED) model of \ngc from \citet{Meh15a}, and the models of \citet{Kaas14} for the warm absorber, the X-ray emission lines, and the X-ray obscurer, to fit the \swift data. This modeling with the {\tt SPEX} package \citep{Kaa96,Kaas20} uses the {\tt xabs} absorption model \citep{dePl04,Stee05}, which calculates both the photoionization solution and the X-ray spectrum, as described in \citet{Mehd16a} for modeling the \swift data. The X-ray covering fraction (\cx) is one of the parameters of the {\tt xabs} model. 

The total column density \NH of the warm absorber in the 2002 unobscured epoch is ${1.2 \times 10^{22}}$~\cm \citep{Kaas14}. In the obscured epoch (i.e. 2012 onwards) new additional absorption by the obscurer with a total column density \NH of ${1.1 \times 10^{23}}$~\cm appears \citep{Kaas14}. The spectroscopy and timing analysis of \citet{Mehd16a} examined the variability of different spectral components and their parameters in \ngc. It was found that the variability of the X-ray hardness ratio $R$ is predominantly caused by changes in \cx of the obscurer providing good fits to all obscured data. Compared to the dominant \cx variability, signature of any \NH variability is relatively too small for \swift measurements. Thus, adopting a constant obscurer \NH during the obscured epoch (2012 onwards), a relation expressing \cx as a function of $R$ was derived in \citet{Mehd16a}: ${\cx = 0.46 + 1.34\, R - 0.91\, R^{2}}$. We use this relation to obtain \cx of the obscurer for \swift data up to 2022.

In the $R$ lightcurve of Figure \ref{fig_lc}, the corresponding \cx of the obscurer is shown on the right axis, which applies to the 2012--2022 obscured data. The earlier 2005 and 2007 \swift data are not obscured, but rather their X-ray continuum is intrinsically fainter and harder \citep{Mehd16a,Detm08}. We verified that the $R$-\cx relation is applicable to the recent \swift data since \cx that is measured from fitting the 2022 \swift spectrum (Figure \ref{fig_spec}) matches the one given by the relation. Thus, the $R$-\cx relation is tried and tested for ${0.5 < \cx < 1}$.

%
\begin{figure}
\centering
\resizebox{0.96\hsize}{!}{\includegraphics[angle=0]{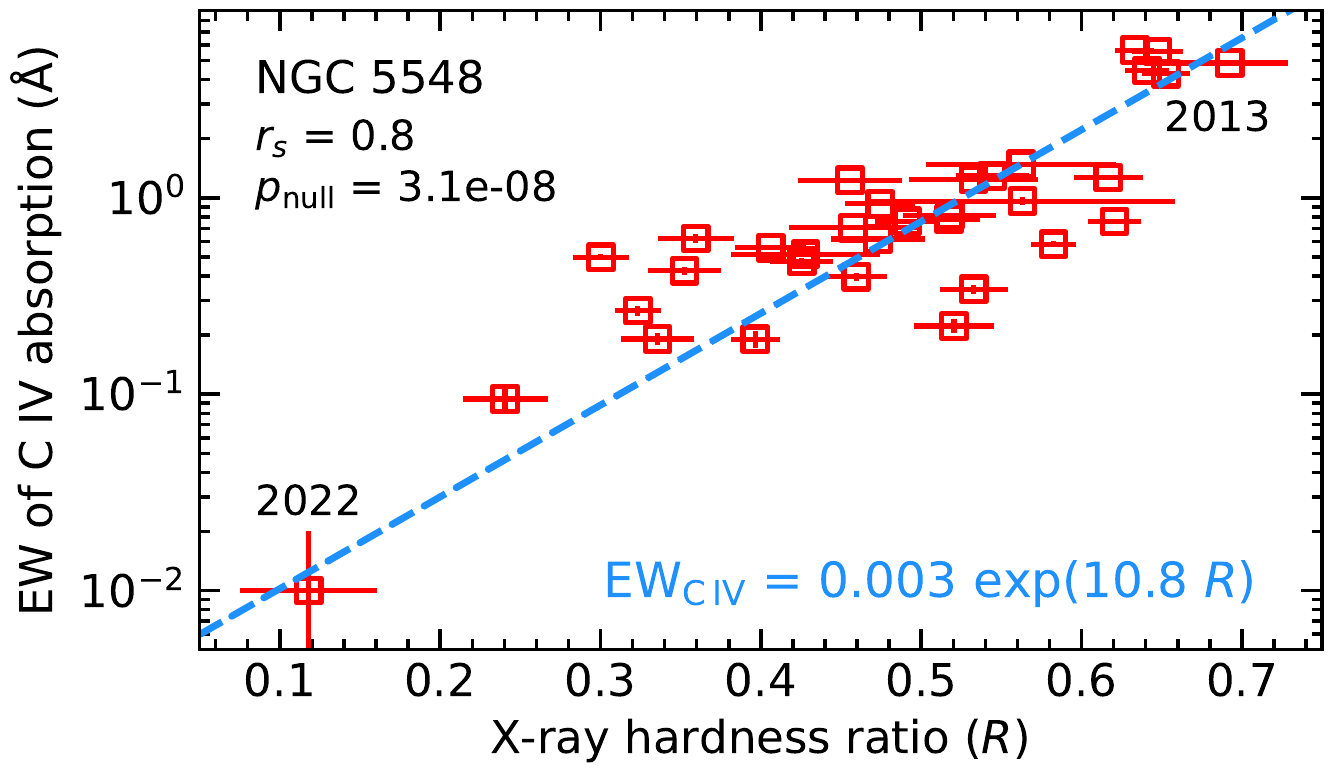}}
\resizebox{0.96\hsize}{!}{\includegraphics[angle=0]{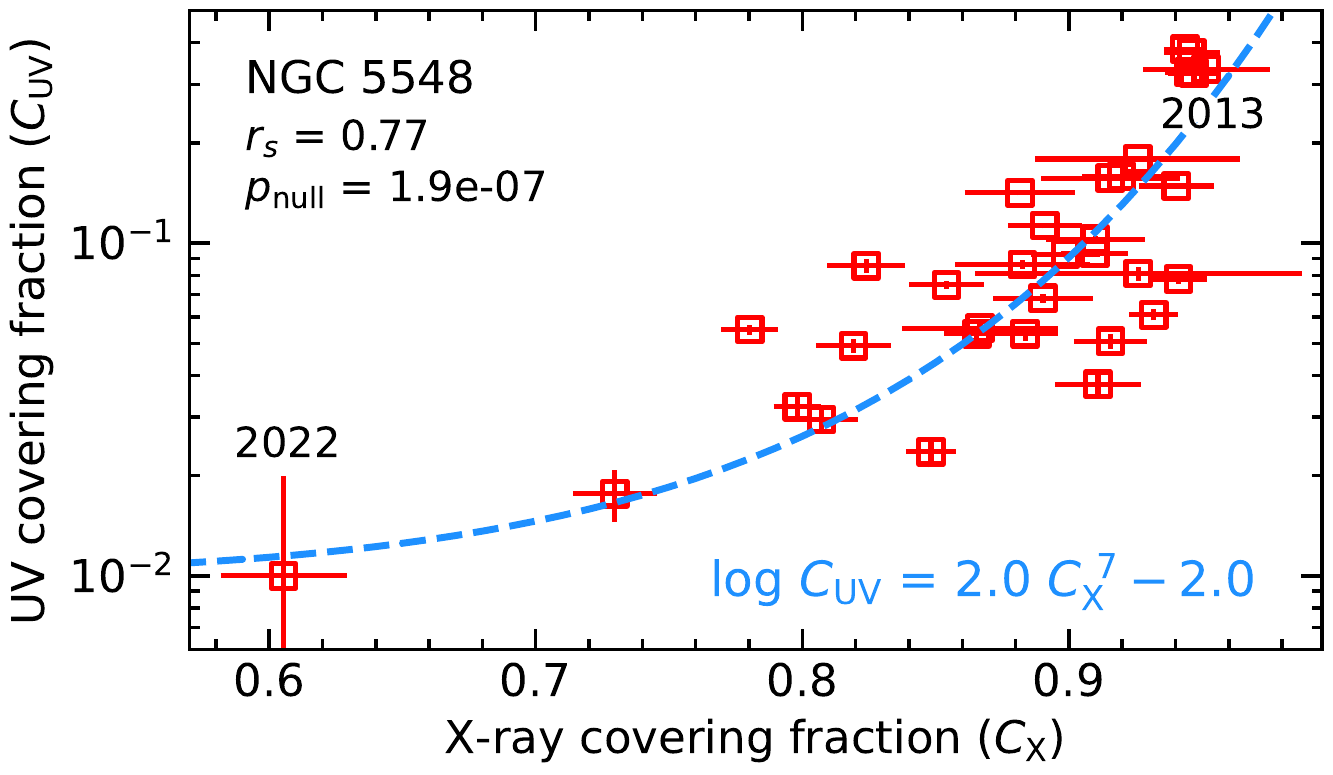}}
\caption{Top panel: relation between the equivalent width (EW) of the broad \ion{C}{4} UV absorption feature and the X-ray hardness ratio ($R$). Bottom panel: relation between the \civ UV covering fraction (\cuv) and the X-ray covering fraction (\cx) of the obscuring wind. Data correspond to contemporaneous \swift and HST/COS observations, taken in 2013, 2014, 2016, and 2022. A time bin size of one week is used for averaging the data. The Spearman’s rank correlation coefficient ($r_s$) and the null hypothesis probability ($p_{\rm null}$) are given in the inset, showing both relations are statistically highly significant. The best-fit function describing each relation is given in the inset and displayed as a dashed line. 
\label{fig_relation}}
\vspace{0.1cm}
\end{figure}

The new HST/COS spectra were analyzed following the spectral modeling and method of \citet{Kris19b}, where all HST/COS data prior to our 2022 observations are presented, including the lightcurve of the broad \civ absorber. The model consists of multiple Gaussian components for fitting the profile of each emission line, namely \civ and \lya. The broad absorption feature on the blue side of the emission line is fitted with Gaussian absorption components, accounting for the blending of the \civ doublet lines. The shift, width, and the line ratios of the components are tied together in a physically-consistent fashion. These broad UV absorption lines from optically-thick gas are saturated. However, they do not appear as ``black'' lines because they partially cover the underlying UV radiation. The spectral fitting of the line profiles, described fully in \citet{Kris19b}, yields the equivalent width (EW) and the corresponding covering fraction of the UV absorber, which is assumed to apply uniformly to the line and continuum emission. In Figure \ref{fig_relation} the EW of the broad \civ absorber (\ewciv) and its covering fraction (\cuv) are plotted versus the X-ray parameters $R$ (top panel) and \cx (bottom panel). These data correspond to when contemporaneous \hst/COS and \swift observations are available, which are taken in 2013, 2014, 2016, and 2022. Compared to the extended long-term monitoring with \swift (Figure \ref{fig_lc}), there is less coverage with \hst over the years. A time bin size of one week is used for averaging the displayed data in Figure \ref{fig_relation}. Interestingly, we find highly significant correlations between the UV absorption and the X-ray obscuration parameters, which we discuss and interpret below.

\section{Discussion and conclusions} 
\label{sect_discus}

\ngc is a key target for ascertaining the general characteristics of obscuring winds in AGN. The long-term \swift monitoring of \ngc shows that its X-ray hardness ratio $R$ has gradually declined over the years, reaching a minimum in 2022 (Figure \ref{fig_lc}). Our study of the corresponding X-ray spectra finds that the long-term $R$ decline is caused by reduction in the X-ray absorption of the obscuring wind. The new \swift and \hst/COS spectra (Figure \ref{fig_spec}) show that as the X-ray obscuration diminishes, the broad \civ absorber nearly vanishes, affirming their physical connection. This is while the intrinsic continuum (UV and hard X-rays) in 2013 and 2022 are vary similar (Figure \ref{fig_spec}). Furthermore, the fact that the narrow \ion{C}{2} absorption lines are diminished in the 2022 COS spectrum (Figure \ref{fig_spec}, bottom panel) demonstrates that the obscurer no longer extensively shields the warm absorber from the ionizing Lyman continuum.

Our long-term variability study of \ngc reveals highly-significant correlations between the broad \civ absorber and the parameters of the X-ray obscurer (Figure \ref{fig_relation}). We note that these UV/X-ray correlations, probing 2013--2022, remain statistically significant regardless of the time bin size. In Figure \ref{fig_relation} the relation between \ewciv and $R$ is best represented by an exponential function, while a power-law best describes $\log \cuv$ as a function of \cx. The functions are provided inset in the panels of Figure \ref{fig_relation}. These empirical relations between the UV and X-ray parameters are not simple linear relations because the observed parameters are the culmination of multiple inextricable aspects, such as the sizes of the X-ray and UV sources (in both the disk and the BLR), and the different line-of-sights and their clumpiness towards the X-ray and UV sources. In other words, because of inhomogeneities in the wind and size differences of the UV and X-ray sources, the local properties of the gas responsible for the UV absorption are not necessarily identical to that responsible for the X-ray absorption. Nonetheless, the UV and X-ray parameters increase/decrease together. Our discovered \ewciv-$R$ and \cuv-\cx relations in \ngc support the premise that X-ray obscuration and UV broad absorption lines (BALs) in AGN are generally interconnected.

Apart from \civ, the broad \lya absorption line also shows a correlation with the X-ray obscuration, but with smaller statistical significance (${r_{s}=0.42}$ and ${p_{\rm null} = 0.02}$). We find that this weaker \lya correlation results from the presence of a relatively constant component of the broad \lya absorber in our line of sight. Thus, while in 2022 the \civ absorber nearly disappears, the \lya absorber, as well as the X-ray obscuration, remain present. We explain this difference between \civ and \lya through the clumpy nature of the wind, where higher-density (lower-ionization) zones are embedded in a lower-density (higher-ionization) medium. Because \lya is more abundant than \civ and can originate from a wider range of ionizations, when the \civ absorption zone dissipates the more expansive lower-density medium still produces \lya and X-ray absorption, albeit at a diminished level. Similar behavior has been observed in the high-ionization absorption troughs of BAL quasar SBS~1542+541, where the covering fractions of different broad absorption lines increase with increasing ionization stage \citep{Telf98}. Furthermore, the clumpiness of such disk winds is supported by hydrodynamical simulations, which find that disk winds can become fragmented due to thermal instabilities as a result of intense X-ray irradiation (see e.g. \citealt{Wate22}).

Our long-term variability study of \ngc shows that \cx of the obscuring wind declined from about 0.95 in 2013 to as low as 0.5 in 2022, while \cuv dropped from 0.3 in 2013 to almost zero in 2022. Interestingly, the fact that the UV absorber nearly vanishes while X-ray obscuration is still substantial may explain why in some obscured Seyfert-1 galaxies no broad UV absorption counterpart was detected. In NGC~3227 \citep{Mehd21,Mao22} obscuration with ${\cx \sim 0.6}$ was found without any significant broad UV absorption. In \ngc when \cx is 0.6, \cuv is effectively negligible; see the \cuv-\cx relation (Figure \ref{fig_relation}, bottom panel). At such \cuv the broad UV absorber would have not been apparent in the lower signal-to-noise spectrum of NGC~3227. Because the UV-radiating sources (disk and BLR) are much more extended than the compact X-ray source, a partially X-ray covering absorber may cover too little of the UV sources to produce detectable UV absorption.

A major difference between obscuration in \ngc and in other Seyfert-1 galaxies, such as NGC~3783 and NGC~3227, has been the duration of the obscuration. In those AGN the transient obscuration lasts weeks/months, however, in \ngc it has been continuously present for a decade, since at least February 2012 (Figure \ref{fig_lc}). This is likely because the obscuring wind in \ngc is a much larger structure surrounding the disk \citep{Dehg19b}. On short timescales (< few days) there are occasional blips when the X-ray obscuration and the broad UV absorption sharply diminish in \ngc \citep{Kris19b}, however, they resume shortly after as inhomogeneities in the wind traverse in our line of sight. On the long timescales that we have explored in this paper we see that the obscuration has been gradually evolving over the years, reaching a peak in strength in 2013--2017 and since then declining (Figure \ref{fig_lc}). 

Our studies of the intrinsic continuum in \ngc \citep{Meh15a,Mehd16a}, as well as NGC~3783 \citep{Mehd17} and NGC~3227 \citep{Mehd21}, show that during the obscuration epochs changes in the continuum parameters and the luminosities are moderate and typical of a Seyfert-1 galaxy. This is in contrast to some changing-look AGN, such as NGC~3516 \citep{Mehd22a}, where the optical/UV emission from the disk and the X-ray emission from the corona have both undergone major transformations. Thus, spectral changes in the obscured AGN and the changing-look AGN have different characteristics. Interestingly, in \ngc from 2000 to 2012 the UV and optical continuum, as well as the BLR emission lines, underwent significant dimming \citep{Cren03,Kris19b}. This was prior to the start of the obscuration epoch in 2012. \citet{Kris19b} suggest that the increase in the accretion rate associated with recovery from that earlier period re-inflated the BLR and triggered the obscuring outflow in \ngc. Since then, however, there are no significant changes in the accretion or coronal activity (\citealt{Meh15a,Mehd16a}, and this work). In the recent 2022 epoch the intrinsic UV and hard X-ray continuum are at similar levels to the 2013 epoch (Figure \ref{fig_spec}), implying similar intrinsic SEDs and luminosities. Thus, the long-term evolution of the obscuring disk wind since 2012 is not connected to any apparent accretion or coronal changes.

%
\begin{figure}
\centering
\resizebox{0.95\hsize}{!}{\includegraphics[angle=0]{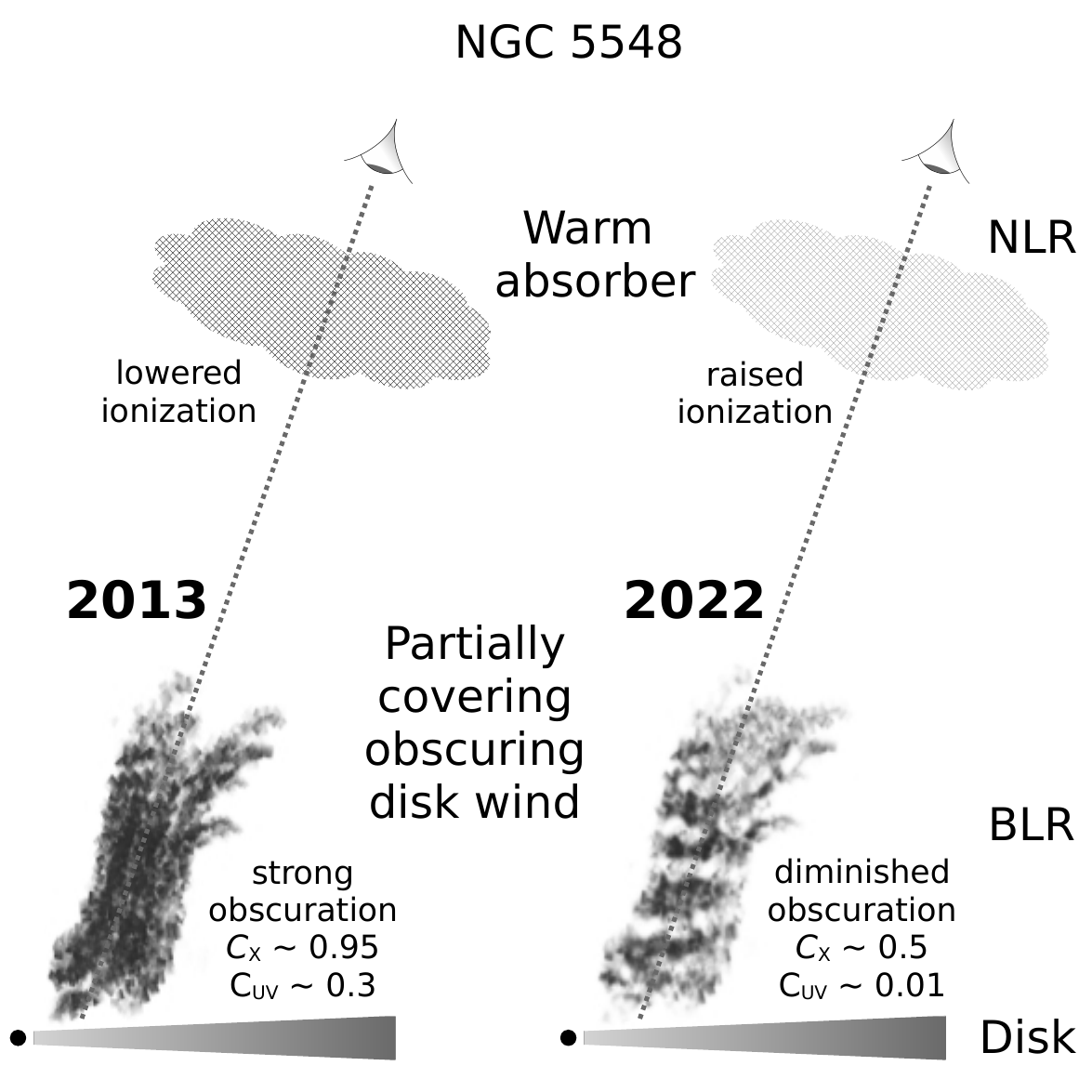}}
\caption{Illustration of the long-term transformation of the obscuring disk wind in \ngc. The outflowing streams from the disk have become more porous and permeable in 2022, hence the observed decline in the X-ray and UV covering fractions (\cx and \cuv). The originally continuous gas supply from the disk in 2013 has gradually become more intermittently powered by the disk.
\label{fig_cartoon}}
\vspace{-0.0cm}
\end{figure}

We illustrate how the obscuring wind in \ngc may have evolved in Figure \ref{fig_cartoon}. Compared to the state of strong obscuration in 2013, the obscurer in 2022 is more porous and permeable. This would explain the significantly smaller \cx and \cuv that we observe in 2022. While in 2013 the supply of gas outflow from the disk appeared to be continuous and uninterrupted, it may have gradually become intermittent, thereby increasing gaps and interstices in the clumpy structure of the wind. This in turn may be due to the mechanism powering the disk wind not being as efficient as it used to be, and perhaps eventually the obscuring wind may fully vanish. There are however no discernible long-term changes in the accretion or coronal activity of \ngc since 2012. 

Irradiation can play an important role in the fragmentation of the wind and hence changing its observed covering fractions. It can induce thermal instability \citep{Wate22} as well as raise the ionization of the lower-density interstices that grow over time. However, irradiation is likely not the primary driver of the long-term decline in the obscuration as there are no corresponding long-term changes in the ionizing SED of \ngc. The obscuring winds in AGN are found to be generally transient and episodic in nature. For example, in NGC~3783 the obscuration, lasting for weeks/months, episodically recurs \citep{Kaas18}. Also, in \ngc there are short intervals where obscuration almost disappears without being a response to the ionizing SED \citep{Mehd16a,Kris19b}. Thus, such intermittent and episodic variations in the disk wind would enhance its clumpiness and partial covering over time. Long-term \swift and \hst monitoring of different AGN that undergo transient obscuration events would be useful for advancing our understanding of the driving mechanism of these disk winds.

\vspace{-0.2cm}
\begin{acknowledgments}
\noindent This work was supported by NASA through a grant for HST program number 16842 from the Space Telescope Science Institute, which is operated by the Association of Universities for Research in Astronomy, Incorporated, under NASA contract NAS5-26555. We acknowledge the use of public Swift data in this work, supplied by the UK Swift Science Data Centre at the University of Leicester. This research has made use of archival Chandra/LETG data. We thank the anonymous referee for providing constructive comments and suggestions that improved the paper.
\end{acknowledgments}
\vspace{-0.2cm}
\facilities{HST (COS), Swift, Chandra (LETG)}
\vspace{-0.4cm}
\bibliographystyle{aasjournal}
\bibliography{references}{}
\end{document}